\documentclass{article}
\begin{document}
\hoffset = -2 truecm \voffset = -2 truecm
\newcommand{\dfrac}[2]{\frac{\displaystyle #1}{\displaystyle #2}}
\title{Particle density in zero temperature symmetry restoring phase transitions
in four-fermion interaction models\footnote{This work was supported by the National
Natural Science Foundation of China.} \\
}
\author{Zhou Bang-Rong  \\
Department of Physics, Graduate School of the \\
Chinese Academy of Sciences, Beijing 100039, China \\
and CCAST (World Lab.),P.O. 8730, Beijing 100080, China }
\date{}
\maketitle
\begin{abstract}
By means of critical behaviors of the dynamical fermion mass in four-fermion
interaction models, we have shown by explicit calculations that when $T=0$ the
particle density will have a discontinuous jumping across the critical chemical
potential $\mu_c$ in $2D$ and $3D$ Gross-Neveu (GN) model and these physically
explain the first order feature of corresponding symmetry restoring phase
transitions. For second order phase transitions in $3D$ GN model when $T\rightarrow
0$ and in $4D$ Nambu-Jona-Lasinio (NJL) model when $T=0$, it has been proven that
the particle density itself will be continuous across $\mu_c$ but its derivative
over the chemical potential $\mu$ will have a discontinuous jumping. The results
give a physical explanation of implications of the tricritical point
$(T,\mu)=(0,\mu_c)$ in 3D GN model. The discussions also show effectiveness of the
critical analysis approach of phase transitions.
\end{abstract}
\emph{PACS numbers}: 11.10.Wx; 12.40.-y;11.30.Qc \\
\emph{Key words}: Gross-Neveu and Nambu-Jona-Lasinio model, symmetry restoration at
zero temperature and high density, particle number density, first and second order
phase transition  \\  \\  \\
\section{Introduction}\label{Intro}
Four-fermion interaction models \cite{kn:1,kn:2} are good laboratories to research
dynamical symmetry restoring phase transitions at high temperature and high density
\cite{kn:3,kn:4,kn:5}. It has been proven that when temperature $T$ goes to zero,
the feature of the phase transitions in these models strongly depend on dimension
$D$ of time-space \cite{kn:6,kn:7,kn:8}. It is shown that that the symmetry
restoring phase transitions at high fermion chemical potential $\mu$ and $T=0$ are
first order when $D=2$ and $D=3$ however second order when $D=4$ and the momentum
cutoff of the loop integrations is large enough. The above conclusions can come from
analyses of critical behaviors of the dynamical fermion mass $m$ as the order
parameter of symmetry breaking. The critical behaviors of the order parameter could
show the essential characteristics, however, they do not directly explain the
physical realization of the above phase transitions. The latter is just what we want
to explore in this paper. Since particle number density is a thermodynamical
quantity directly relative to the chemical potential $\mu$, we will calculate it in
variant cases and combine the results with the critical behaviors of the order
parameter obtained in Ref.\cite{kn:7,kn:8} so as to further expound physical
characteristics of the above first and second order phase transitions. \\
\indent The total particle number density $n_D(T,\mu,m)$ in $D$ dimensional
time-space should be a difference between fermion and antifermion density and at
finite $T$ and finite $\mu$ it can be expressed by \cite{kn:9}
\begin{eqnarray}
n_D(T,\mu,m)&=&\int_{-\infty}^{\infty}\dfrac{d^{D-1}\vec{p}}{(2\pi)^{D-1}}
\left[\dfrac{1}{e^{\beta(\sqrt{\vec{p}^2+m^2}-\mu)}+1}-(\mu\rightarrow -\mu)\right],
\ \beta=1/T
\nonumber \\
   &=&\frac{2}{\Gamma(\frac{D-1}{2})(4\pi)^{(D-1)/2}}\int_0^{\infty} d p p^{D-2}
   \left[\dfrac{1}{e^{\beta(\sqrt{p^2+m^2}-\mu)}+1}-(\mu\rightarrow -\mu)\right]\\
   &=& \frac{2T^{D-1}}{\Gamma(\frac{D-1}{2})(4\pi)^{(D-1)/2}}\int_0^{\infty}dx
   \left[\dfrac{x^{D-2}}{e^{\sqrt{x^2+y^2}-\alpha y}+1}-(\alpha\rightarrow -\alpha)\right]
   , y=\beta m, \alpha=\mu/m,
\end{eqnarray}
where $m$ is the fermion mass and in a four-fermion interaction model with
spontaneous symmetry breaking, it will be replaced by the dynamical fermion mass
$m\equiv m(T,\mu)$ caused by the bilinear fermion condensates \cite{kn:7}. We will
discuss the cases of $D=2,\ 3$ and $4$ respectively in Sects. \ref{D=2}, \ref{D=3}
and  \ref{D=4}, then come to our conclusions in Sect. \ref{concl}.
\section{D=2}\label{D=2}
First let us consider the case of $m=0$. We may find out from Eq.(1) that
\begin{equation}
n_2(T,\mu,m=0)=\frac{1}{\pi}\int_0^{\infty}dp\left[\frac{1}{e^{\beta(p-\mu)}+1}-(\mu\rightarrow
-\mu)\right]=\frac{T}{\pi}\left[\ln(1+e^{\beta\mu})-\ln(1+e^{-\beta\mu})\right]=\frac{\mu}{\pi}.
\end{equation}
Eq.(3) represents the density of massless fermions or the particle density in the
case without spontaneous symmetry breaking. Next turn to the case of $m\neq 0$. Let
$z=\sqrt{x^2/y^2+1}$, we may get from Eq.(2) that
\begin{equation}
n_2(T,\mu,m)=\frac{m}{\pi}\int_1^{\infty}dz\frac{z}{\sqrt{z^2-1}}
\left[\frac{1}{e^{y(z-\alpha)}+1}-(\alpha\rightarrow -\alpha)\right]
\end{equation}
We will concern ourselves only with the results in the limit $T\rightarrow 0$ (or
$y\rightarrow \infty$). It is easy to see that when $y\rightarrow \infty$, the
second term in Eq.(4) becomes zero and the non-zero contribution comes only from the
first term when $\alpha >1$, owing to the interval of the integral variable $z$ is
from 1 to $\infty$.  Thus we may obtain from Eq.(4) that
\begin{eqnarray}
n_2(T=0,\mu,m) &=& \left\{\matrix{
                            0 & {\rm when} \ \alpha\leq 1 \cr
                             \dfrac{m}{\pi}\int_1^{\alpha}dz\dfrac{z}{\sqrt{z^2-1}}
                             &{\rm when} \ \alpha > 1 \cr}\right.
                             \nonumber\\\vspace{0.5cm}
  &=&\dfrac{\theta(\mu-m)}{\pi}\sqrt{\mu^2-m^2}
\end{eqnarray}
Eq.(5) is applicable to the general case of free fermions with $m\neq 0$. Obviously,
$n_2(T=0,\mu,m)$ usually varies continuously when $\mu$ crosses over $m$. However,
in a four-fermion interaction model with spontaneous symmetry breaking and
restoration, there could be different case. It has been proven that \cite{kn:7}, in
a $2D$ GN model, when $T=0$ the order parameter $m$ will varies in the following
form:
\begin{equation}
m(T=0,\mu)=m(0)\theta[m(0)-\mu]
\end{equation}
where $m(0)$ is the dynamical fermion mass at $T=\mu=0$ and just the critical
chemical potential $\mu_c$ at $T=0$. Substituting Eq.(6) into Eq.(5), we will reach
the conclusion that
\begin{equation}
n_2(T=0,\mu,m)=\left\{\matrix{
                       0 & {\rm when} \ \mu\leq m(0) \cr
                       \mu/\pi & {\rm when} \ \mu>m(0) \cr}\right..
\end{equation}
Eq.(7) indicates that the jumping of the order parameter $m$ from $m(0)$ to $0$
across over $\mu_c=m(0)$ leads to jumping of the thermodynamical particle density
$n_2(T=0,\mu,m)$ from $0$ to $\mu/m$, i.e. from the zero value with spontaneous
symmetry breaking to the density of free fermion with $m=0$ after the symmetries are
restored.  This clearly shows physical characteristic of the first order phase
transition at $T=0$ and $\mu=\mu_c$ in $2D$ GN model.
\section{D=3}\label{D=3}
In this case, we will first obtain from Eq.(2) that
\begin{eqnarray}
n_3(T,\mu,m)&=& \frac{T^2}{2\pi}\int_0^{\infty}dx
   \left[\frac{x}{e^{\sqrt{x^2+y^2}-\alpha y}+1}-(\alpha\rightarrow -\alpha)\right] \nonumber\\
   &=&\frac{m^2}{2\pi}\int_1^{\infty}dz z
   \left[\frac{1}{e^{yz-r}+1}-(r\rightarrow -r)\right], r=\beta \mu \nonumber \\
   &=& -\frac{m^2}{2\pi y}\int_1^{\infty}dz z \frac{\partial}{\partial z}
   \left[\ln(1+e^{-yz+r})-\ln(1+e^{-yz-r})\right] \nonumber \\
  &=& \frac{Tm}{2\pi}\ln\frac{1+e^{-y+r}}{1+e^{-y-r}}+\frac{T^2}{\pi}\sum_{k=1}^{\infty}
 \frac{(-1)^{k+1}}{k^2}\sinh (k r) \ e^{-ky} \\
&=&
\frac{Tm}{2\pi}\ln\frac{1+e^{(\mu-m)/T}}{1+e^{-(\mu+m)/T}}+\frac{T^2}{2\pi}\sum_{k=1}^{\infty}
 \frac{(-1)^{k+1}}{k^2} \left[e^{k(\mu-m)/T}-e^{-k(\mu+m)/T}\right]
\end{eqnarray}
Eq.(9) will lead to the $T\rightarrow 0$ limit of $n_3(T,\mu,m)$
\begin{equation}
n_3(T=0,\mu,m)=\frac{1}{4\pi}\theta(\mu-m)(\mu^2-m^2).
\end{equation}
Assuming that $y=m/T \ll 1$ (e.g. near a critical point of a second order phase
transition), then we can expand $n_3(T,\mu,m)$ expressed by Eq.(8) in power of $y$,
keeping only the terms up to the order of $y^2$, and obtain
\begin{equation}
n_3(T,\mu,m)\simeq n_3(T,\mu,m=0) -\frac{m^2}{4\pi}\frac{e^r-1}{e^r+1}
\end{equation}
where
\begin{eqnarray}
n_3(T,\mu,m=0)&=&\frac{T^2}{\pi}\sum_{k=1}^{\infty}
 \frac{(-1)^{k+1}}{k^2}\sinh (k\beta \mu) \nonumber \\
 &=&\frac{\mu^2}{4\pi}+\frac{T^2\pi}{12}
 -\frac{T^2}{\pi}\sum_{k=1}^{\infty}
 \frac{(-1)^{k+1}}{k^2}e^{-k\beta \mu}.
\end{eqnarray}
Noting that, different from the case of $D=2$, the total particle density is
temperature-dependent and when $T=0$ we have
\begin{equation}
n_3(T=0,\mu,m=0)=\mu^2/4\pi.
\end{equation}
Eq.(10) is a general expression for free fermions with $m\neq 0$ at $T=0$. However,
for $3D$ GN model with symmetry restoring phase transition when $T=0$, we have the
order parameter $m$ whose critical behavior is similarly expressed by Eq.(6)
\cite{kn:7}.  Hence we obtain
\begin{equation}
n_3(T=0,\mu,m)=\left\{\matrix{
                       0 & {\rm when} \ \mu\leq m(0) \cr
                       \mu^2/4\pi& {\rm when} \ \mu>m(0) \cr}\right.,
\end{equation}
i.e.the zero temperature particle density $n_3(T=0,\mu,m)$ will jump from $0$ with
spontaneous symmetry breaking to $\mu^2/4\pi$ of massless fermions (symmetries being
restored) when $\mu$ crosses over the critical point $m(0)$ of the phase transition.
Such jumping of the particle density definitely indicates the first order feature of
the phase transition.\\
\indent On the other hand, when $T\neq 0$, near a critical chemical potential
$\mu_c$, the order parameter $m$ will continuously vary in the form \cite{kn:7}
\begin{equation}
m^2\simeq 2T\sinh\frac{\mu}{T} \;\; (\mu_c-\mu),  \ {\rm when} \ T\neq 0.
\end{equation}
Substituting Eq.(15) into Eq.(11) we will obtain
\begin{equation}
n_3(T,\mu,m)=n_3(T,\mu,m=0)-\left\{\matrix{
                    \dfrac{T}{2\pi}\dfrac{e^r-1}{e^r+1}\sinh\dfrac{\mu}{T}\;\;
                    (\mu_c-\mu) & {\rm when} \ \mu \stackrel{<}{\sim} \mu_c \cr
                    0&{\rm when} \ \mu > \mu_c \cr}
                    \right.
\end{equation}
and
\begin{equation}
\frac{\partial n_3(T,\mu,m)}{\partial \mu}\simeq \frac{\partial
n_3(T,\mu,m=0)}{\partial
    \mu}+\left\{\matrix{
    \dfrac{T}{2\pi}\dfrac{e^r-1}{e^r+1}\sinh\dfrac{\mu}{T}
    &{\rm when} \ \mu \stackrel{<}{\sim} \mu_c \cr
    0&{\rm when} \ \mu > \mu_c \cr}
                    \right..
\end{equation}
It is seen from Eqs.(16) and (17) that at $T\neq 0$ when $\mu$ crosses over $\mu_c$,
$n_3(T,\mu,m)$ continuously increases from the value less than $n_3(T,\mu, m=0)$ up
to $n_3(T,\mu, m=0)$, however the derivative $\partial n_3(T,\mu,m)/\partial \mu$
has a discontinuous jumping at $\mu=\mu_c$. This shows the second order of the phase
transition. We emphasize that this feature can be maintained when $T$ is finite and
very small as long as $\mu_c-\mu$ is also very small so that $m^2/T^2\ll 1$ is kept.
The limiting point will be $(T=0,\mu=m(0))$ which is the same as the first order
phase transition point when $T=0$. The above discussions give a physical explanation
of implications of the tricritical point $(T,\mu)=(0,m(0))$ in $3D$ GN model: The
particle density will jump from $0$ to $\mu^2/4\pi$ when the point $(0,m(0))$ is
approached along the $(T=0,\mu)$ axis but will continuously increase from a lower
value to $\mu^2/4\pi$ when the point is approached along the $T-\mu$
critical curve of the second order phase transitions. \\
\section{D=4}\label{D=4}
For $4D$ NJL model, when $m=0$ we will have
\begin{equation}
n_4(T,\mu,m=0)=\frac{1}{2\pi^2}\int_0^{\infty}dp
p^2\left[\frac{1}{e^{\beta(p-\mu)}+1}-(\mu\rightarrow -\mu)\right]
\end{equation}
The equivalent differential equation is
\begin{equation}
\frac{\partial^2n_4(T,\mu,0)}{\partial \mu^2}=\frac{T}{\pi^2}\left[
\ln(1+e^{\beta\mu})-\ln(1+e^{-\beta\mu}) \right]=\frac{\mu}{\pi^2}
\end{equation}
with the boundary conditions
\begin{equation}
n_4(T,\mu=0,0)=0, \ \left.\frac{\partial n_4(T,\mu,0)}{\partial
\mu}\right|_{\mu=0}=\frac{T^2}{6}.
\end{equation}
The solution of Eq.(19) satisfying the conditions in Eq.(20) is that
\begin{equation}
n_4(T,\mu,0)=\frac{\mu^3}{6\pi^2}+\frac{T^2}{6}\mu\stackrel{T\rightarrow
0}{=}\frac{\mu^3}{6\pi^2}.
\end{equation}
When $m\neq 0$ and $T\rightarrow 0\ (y\rightarrow \infty)$, we may obtain from
Eq.(2) that
\begin{eqnarray}
n_4(T=0,\mu,m)&=&\lim\limits_{y\rightarrow
\infty}\frac{m^3}{2\pi^2}\int_1^{\infty}dz z
\sqrt{z^2-1}\left[\frac{1}{e^{y(z-\alpha)}+1}-(\alpha\rightarrow -\alpha)\right]\nonumber \\
   &=&\left\{\matrix{0,& {\rm when} \ \alpha\leq 1 \cr
                     \dfrac{m^3}{2\pi^2}\int_1^{\alpha}dz z \sqrt{z^2-1}, &{\rm
                     when} \ \alpha>1 \cr}\right. \nonumber \\
   &=&\theta(\mu-m)\frac{\mu^3}{6\pi^2}\left(1-\frac{m^2}{\mu^2}\right)^{3/2}
\end{eqnarray}
Eq.(22) is a general result of free fermions and usually does not involve any
spontaneous symmetry breaking and restoration. Now consider the case in which
symmetry restoring phase transition is assumed to occur. Thus we should identify $m$
in Eq.(22) with the dynamical fermion mass. It has been proven that in a $4D$ NJL
model, the dynamical fermion mass $m$ the following behavior \cite{kn:8}
\begin{equation}
\matrix{m=m(0), &{\rm when} \ \mu\leq m(0) \cr
\mu_{c0}^2-\mu\sqrt{\mu^2-m^2}=\dfrac{m^2}{2}\ln\dfrac{\Lambda^2+m^2}{(\mu+\sqrt{\mu^2-m^2})^2},
&{\rm when} \ m(0)<\mu<\mu_{c0} \cr}
\end{equation}
where $\Lambda$ is the 4 dimensional Euclidean momentum cutoff and the critical
chemical potential $\mu_{c0}$ is defined by
\begin{equation}
\mu_{c0}^2=\frac{1}{2}m^2(0)\ln\left[\frac{\Lambda^2}{m^2(0)}+1\right].
\end{equation}
We will assume that
\begin{equation}
\mu_{c0}^2\leq \frac{\Lambda^2}{4e}
\end{equation}
so that all the phase transitions will be second order \cite{kn:8}. Combining
Eq.(22) with Eq.(23) we may find that when $\mu\leq m(0)$, $n_4(T=0,\mu,m)=0$ and
when $m(0)<\mu<\mu_{c0}$, $n_4(T=0,\mu,m)$ will increase in the form
$\mu^3(1-m^2/\mu^2)^{3/2}/6\pi^2$ (noting that $\partial m/\partial \mu<0$ in this
region). As soon as $\mu$ is arriving at $\mu_c$ where symmetries will be restored,
$m$ will reduce to zero and $n_4(T=0,\mu,m)$ becomes $\mu^3/6\pi^2$ which is just
the number density of massless fermions given by Eq.(21). It should be emphasized
that, since $m$ reduces to zero at $\mu_c$ continuously, $n_4(T=0,\mu,m)$ changes
from 0 into $\mu^3/6\pi^2$ also in continuous form when $\mu$ varies from $m(0)$ to
$\mu_c$ and this is just the characteristic of a second order phase transition. The
above result can be outlined as
\begin{equation}
n_4(T=0,\mu,m)=\left\{\matrix{
                       \dfrac{1}{6\pi^2}(\mu^2-m^2)^{3/2}& {\rm when} \ m(0)<\mu\leq
                       \mu_{c0} \cr
                       \dfrac{\mu^3}{6\pi^2} &{\rm when} \ \mu >
                       \mu_{c0} \cr}\right.
\end{equation}
We can also prove that the derivative of $n_4(T=0,\mu,m)$ over $\mu$ will be
discontinuous when $\mu$ crosses over $\mu_{c0}$.  To this end, we will use the
critical behavior of the squared order parameter $m^2$:
\begin{equation}
m^2\simeq \frac{(\mu^2_{c0}-\mu^2)}{\left(
\ln\dfrac{\Lambda}{2\mu}-\dfrac{1}{2}\right)}\simeq 2(\mu_{c0}-\mu)\; \mu_{c0} \;
b(\mu_{c0}), \ {\rm when} \ \mu \stackrel{\large <}{\sim} \mu_{c0},
\end{equation}
where $b(\mu_{c0})=1/\left(\ln\dfrac{\Lambda}{2\mu_{c0}}-\dfrac{1}{2}\right) >0$
owing to Eq.(25). Substituting Eq.(27) into Eq.(26), we get
\begin{equation}
n_4(T=0,\mu,m)=\left\{\matrix{
   \dfrac{1}{6\pi^2}[\mu^2-2(\mu_{c0}-\mu) \; \mu_{c0} \; b(\mu_{c0})]^{3/2},& {\rm when} \
   \mu\stackrel{<}{\sim}
                       \mu_{c0} \cr
                       \dfrac{\mu^3}{6\pi^2} &{\rm when} \ \mu>
                       \mu_{c0}\cr}\right.
\end{equation}
and furthermore,
\begin{equation}
\frac{\partial n_4(T=0,\mu,m)}{\partial \mu}=\left\{ \matrix{
             \dfrac{\mu^2}{2\pi^2}+\dfrac{\mu_{c0} \ \mu}{2\pi^2}\;b(\mu_{c0}), &{\rm
             when} \ \mu \stackrel{<}{\sim} \mu_{c0} \cr
             \dfrac{\mu^2}{2\pi^2} &{\rm when} \ \mu>\mu_{c0}\cr}
                      \right..
\end{equation}
Eq.(29) indicates that $\partial n_4(T=0,\mu,m/\partial \mu$ has a jumping over
$\mu=\mu_{c0}$ and is discontinuous. The above results verifies that the discussed
symmetry restoring phase transition in the $4D$ NJL model is second order indeed.\\
\section{Conclusions}\label{concl}
By combining the calculated fermion number densities as a thermodynamical quantity
with the known critical behaviors of the dynamical fermion mass in the four-fermion
interaction models, we have shown that when temperature $T=0$, across over the
critical chemical potential $\mu_c$, the particle number density will have a
discontinuous jumping in $2D$ and $3D$ GN model and these physically explain the
first order feature of corresponding zero temperature symmetry restoring phase
transitions. The second order phase transitions in $3D$ GN model when $T\neq 0$ and
in $4D$ NJL model when $T=0$ and $\mu_{c0}^2\leq \Lambda^2/4e$ are illustrated by
the facts that the densities themselves are continuous but their derivatives over
chemical potential $\mu$ have a discontinuous jumping. These results clearly show
physical difference between the first order and second order symmetry restoring
phase transition in these models, especially give a physical explanation of
implications of the tricritical point in the $3D$ GN model. The whole discussions
also indicate the full effectiveness of critical analysis of the order parameter for
phase transition problem.

\end{document}